# Fiber-based narrowband bright squeezed light generation by double-pass parametric amplification

TIANYI TAN[1], CHANGSHENG YANG[1,2,6], QILAI ZHAO[1], CHENGZI HUANG[1], XIANCHAO GUAN[1,4], ZHONGMIN YANG[1,2,3,4], AND SHANHUI XU[1,2,3,5,7]

[1]State Key Laboratory of Luminescent Materials and Devices and Institute of Optical Communication Materials, South China University of Technology, Guangzhou 510640, China
[2]Guangdong Engineering Technology Research and Development Center of Special Optical Fiber Materials and Devices, Guangzhou 510640, China
[3]Guangdong Provincial Key Laboratory of Fiber Laser Materials and Applied Techniques, South China University of Technology, Guangzhou 510640, China
[4]School of Physics and Optoelectronics, South China University of Technology, Guangzhou 510640, China
[5]Guangdong Engineering Technology Research and Development Center of High-performance Fiber Laser Techniques and Equipments, Zhuhai 519031, China
[6]mscsyang@scut.edu.cn
[7]flxshy@scut.edu.cn

The squeezed states of light become more and more important in the fields of quantum enhanced precision measurement and quantum information. To get this vital continuous variable quantum resource, the generation of squeezed states of light becomes a key factor. In this paper, a compact telecom fiber-based bright squeezed light (BSL) generator is demonstrated. To our knowledge, this is the first time that BSL has been reported in a fiber-based system to date. To obtain the BSL, a double-pass parametric amplifier based on surface-coated lithium niobate waveguide is employed. When the 1550 nm seed laser of the parametric amplifier is blocked, a stable 1.85 dB squeezed vacuum is obtained. With injected seed power of 80 μW, an output power of 18 μW and a squeezing value of 1.04 dB are achieved of the BSL at 1550 nm. Due to the good mode matching in the fiber and the absence of the resonant cavity, this flexible and compact BSL generator has the potential to be useful in out-of-the-laboratory quantum technologies. Moreover, the BSL has a narrow spectral width of 30 kHz, which is inherited from a narrow-linewidth single-frequency seed laser. In addition to being free from the wavelength-dependent losses, the narrowband BSL is also beneficial to improve the signal-to-noise ratio of quantum-enhanced precision measurement.

# 1. INTRODUCTION

The generation of squeezed states of light has been an important technology to prepare continuous variable quantum resources, which have been the foundations of various applications, such as quantum information [1, 2] and quantum computing [3]. In addition to their potential in information technology, squeezed states of light are also active in systems trapped by quantum noise limit to achieve the ultimate accuracy, such as quantum imaging [4, 5], quantum enhanced precision measurement [6-8], and especially gravitational wave detection [9].

In recent years, with the gradually mature development of the squeezing technologies based on the bulk-optics systems, advanced squeezing sources have been reported at various optional wavelengths towards different applications. To date, thanks to the small transmission loss in the free-space, the cavity-enhanced technology and the strict mode matching, the squeezing vacuum sources with the highest squeezing value of 15, 13, and 4 dB are achieved at 1.0, 1.5, and 2.0-μm band, respectively [10-12]. However, the sophisticated structure and huge scale of the squeezing systems limit the popularization of this valuable technology in scientific and technical researches. In response to this difficulty, a series of miniaturized squeezing sources based on fiber or on-chip systems have been implemented [13-16]. Miniaturization of the semi-monolithic optical parametric oscillator is one of the viable methods, by which a 6.2 dB squeezing source has been realized [17, 18]. For further simplification of the squeezing source, the single-pass spontaneous parametric down conversion (SPDC) has become a decent choice for generating the squeezed vacuum. So far, the highest squeezing value achieved in fiber-based system by single-pass SPDC is 3.2 dB [19]. The development of these fiber-based generators of squeezed vacuum has great benefits in improving the accuracy of precision measurements, especially in fiber interferometers [20-22].

The difference from the squeezed vacuum state is that the bright squeezed light (BSL) has a certain coherent amplitude, thus it can be directly used as a signal in some quantum-noise-limited systems, such as quantum imaging [23] and optomechanical magnetometry [24]. On the other hand, the BSL can also be used as the continuous variable quantum entanglement resource in applications such as quantum dense coding and quantum teleportation [25-27]. Generally, without the restriction of the resonator, the squeezed vacuum generated by a single-pass SPDC has a broadband spectrum (tens of nanometers), which is determined by the phase matching bandwidth of the non-linear medium [28]. However, the broadband characteristic means that the applications of such squeezing sources in the systems with wavelength-dependent loss will be limited, such as wavelength division multiplexing systems. In contrast, due to the narrowband spectrum, which is mainly determined by the seed linewidth in optical parametric amplification (OPA), the BSL can solve such problems even without the intervention of the resonant cavity. There are already excellent results of the generation of BSL in bulk-optics systems [29-31]. However, to our best knowledge, the narrowband BSL generated in fiber-based system has been less investigated to date.

In this paper, we report a telecom fiber-based narrowband BSL generated by a double-pass parametric amplification which has a squeezing value of 1.04 dB and a spectral bandwidth of 30 kHz. The whole system employs fiber-based optical components requiring no alignment procedures for spatial mode matching. These advantages guarantee extreme reliability and make our approach a valuable candidate for real-world applications based on BSL.

## 2. EXPERIMENTAL SETUP

The experimental setup is shown in Fig. 1. A 1550 nm single-frequency distributed feedback laser diode (DFB-LD) is employed as the laser source of the system. It has an intensity noise equal to the quantum noise limit in the frequency range greater than 200 kHz and a linewidth of less than 30 kHz. The DFB-LD with 50 mW output is firstly divided by two 50:50 optical couplers (OCs) into local oscillator (LO), seed coherent state, and fundamental laser.

The fundamental laser with a power of ~20 mW is then amplified to 800 mW by a master-oscillator power amplifier (MOPA) with a 3.5-m-long $Er^{3+}/Yb^{3+}$-codoped double-cladding fiber.

The boosted fundamental laser is then injected into the periodically-poled $LiNbO_3$ waveguide 1 (PPLN WG1) for the second harmonic generation (SHG). At a phase-matching temperature of 37.3 °C, the SH laser with a power of 110 mW, which is used as the pump laser of the OPA, is output and transmitted through the polarization-maintaining (PM) single-mode fiber (Nufern 780HP). In order to avoid the influence of residual pump on squeezing value of the BSL, the residual fundamental laser is filtered out by a 4-cm-diameter fiber loop with a segment of 780HP fiber, which does not support the fundamental mode of 1550 nm laser (not represented in Fig. 1) [32]. In addition, the power of the pump injected into PPLN WG2 is controlled by the variable optical attenuator 3 (VOA3) during the experiment.

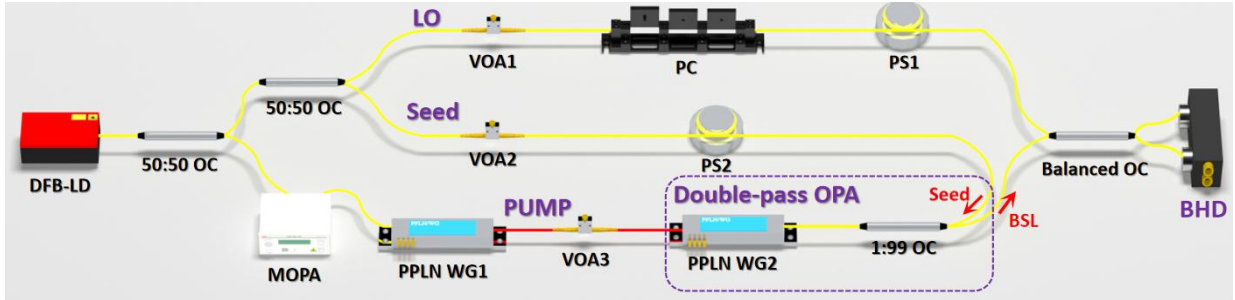

**Fig. 1.** The experimental setup of the narrowband BSL generation. A 1550 nm single-frequency low-noise distributed feedback laser diode (DFB-LD) is divided into local oscillator (LO), seed coherent state (SEED), and fundamental laser by two 50:50 fiber optical coupler (OC). The fundamental laser is boosted by a master-oscillator power amplifier (MOPA) and subsequently injected into the periodically-poled LiNbO$_3$ waveguide 1 (PPLN WG 1). The generated 775 nm laser by second harmonic generation (SHG) is then guided into the PPLN WG 2 as the pump. The SEED is phase controlled by a piezoelectric fiber stretcher (PS2) and then backward guided into the PPLN WG 2 by a 1:99 OC. The seed coherent state interacts with the pump to generate the bright squeezed light (BSL) in the double-pass optical parametric amplification (OPA). The power, polarization and phase of LO are manipulated by a variable optical attenuator (VOA1), a polarization controller (PC) and PS1. Eventually, the BSL signal interferes with the LO on a homemade exquisite balanced OC. And the noise performance is monitor by a balanced homodyne detector (BHD).

The schematic of the double-pass parametric amplification is shown in Fig. 2. The input and output facets of the PPLN WG2 are coated with high-reflective (HR) film corresponding to the seed (1550 nm) and the pump wavelength (775 nm), respectively. And the reflectivity of the HR coating at both facets is greater than 99.99%. The pump laser from SHG is forward guided into the WG2 and reflected back at the output facet. Similarly, the quantum-noise-limited seed coherent state is backward guided into the WG2 by a 1:99 OC and reflected back at the input facet. Meantime, the OPA occurs during the double-pass propagation of the pump and seed laser in WG2. The gain factor of amplification or deamplification depends on the relative phase between two beams which is manipulated by a piezoelectric fiber stretcher (PS). The power of the seed coherent state finally guided into the WG2 is ~80 µW, which is controlled by the VOA2, and the final output power of the BSL signal is about 18 µW.

This double-pass parametric amplification structure doubles the effective nonlinear interactional length in the WG2. In addition, it has other significance. Firstly, it is solved the issue that injection of the seed coherent state from the same fiber as the pump is impossible due to the nonsupport of 1550 nm fundamental mode in the 780HP fiber. Secondly, the HR coating of 1550 nm prevents noisy photons coupling into the parametric amplifier, which may reduce the squeezing factor of the BSL. Finally, the double pass of the pump means that it does not pass through the parametric amplifier, which implies that it is not necessary to insert an additional fiber filter with superfluous insertion loss.

Moreover, to avoid severe phase jitter caused by environmental noise, the entire system except the LO path is passively isolated from sound and vibration. Furthermore, the transmission fibers of the seed, the pump and the BSL are all PM single-mode fiber which insensitive to the environmental noise.

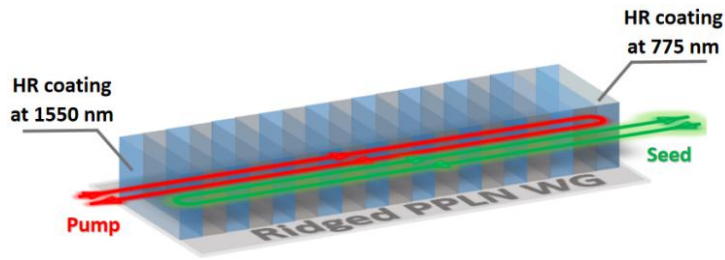

**Fig. 2.** Schematic of the double-pass OPA. The enlarged model of the PPLN WG2 and the propagation paths of the 775 nm pump laser (red) and the 1550 nm seed coherent state (green) in WG2 are shown in this figure. The input and output facets of the WG2 are coated with high-reflective (HR) film corresponding to 1550 nm and 775 nm, respectively.

The noise performance of the generated BSL is detected by a balanced homodyne detector (BHD). The power of LO is adjusted to 5 mW by VOA1, which is 278 times as powerful as that of the BSL. The phase of the LO is scanned by a piezoelectric fiber stretcher (PS) under a 5 Hz triangle-wave modulation. To ensure enough interference efficiency, a polarization controller (PC) is used to adjust the polarization state of the LO to maintain the same of that as the BSL signal. In our experiments, the measured efficiency of interference is greater than 97%. The BSL signal interferes with the LO on a homemade exquisite balanced OC, which has a insertion loss of 0.21 dB and a ratio deviation of less than ±2%. Moreover, the two output arms of the balanced OC are adjusted to the same length. To ensure the squeezing value of the BSL, a pair of photodiodes with the quantum efficiency of greater than 99% is employed in the BHD which includes built-in filters, amplification and differential circuits. The total system transmission efficiency is estimated to be 0.533, which includes waveguide transmission efficiency of 0.75, the fiber transmission efficiency of 0.74, interference efficiency of 0.97, and detection efficiency of 0.99.

## 3. RESULTS AND DISCUSSION

The normalized measured parametric gain of the OPA is shown in Fig. 3. When the pump laser is absent, an output power of ~29 μW is obtained with ~80 μw seed injection due to the loss. Subsequently, considering whole coupling and material losses, with a pump power of 60 mW and a seed laser power of ~80 μW guided in the OPA, the maximum and minimum output power is ~147 μW and ~18 μW, respectively. That means, when the phase of the seed coherent state relative to the pump is varied, the output power is amplified up to 5.06 times and deamplified down to 0.62 times. In order for OPA to work in the deamplification state to generate quadrature amplitude BSL, it is necessary to control the phase shifter to minimize the DC component of the output of the photodetector [33].

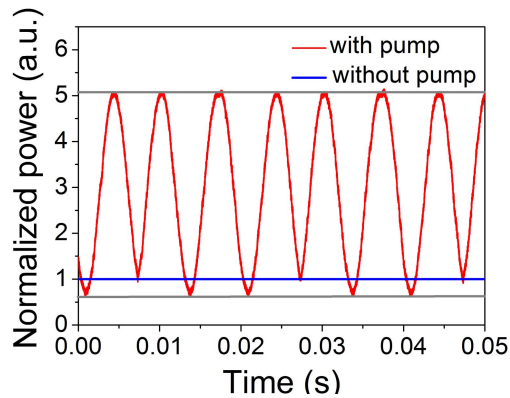

**Fig. 3.** Normalized output power of the OPA versus phase scanning over time with a pump power of 60 mW. The maximum gain of amplification is 5.06 times, and the minimum gain of deamplification is 0.62 times.

The squeezing curves of generated squeezed vacuum and BSL are shown in Fig. 4. (a) and (b), respectively, which are measured by scanning the phase of the LO. When the seed light is blocked by VOA2, a squeezed vacuum with 1.85 dB squeezing and 2.7 dB anti-squeezing is obtained with 60 mW pump. At the same pump power, 80 μw of seed coherent state is injected into the OPA, meanwhile, the PS2 is adjusted to minimize the DC component of the BHD output, which means that the OPA works in the deamplification state. Finally, about 20 µW BSL is obtained with 1.04 dB squeezing and 2.0 dB anti-squeezing. Obviously, when there is no phase locking between the LO and the seed laser, even if passive isolation from environmental noise is taken, the squeezing value of BSL still be affected by the phase jitter. For further development, relative phase locking technology can improve and stabilize this type of generation of BSL.

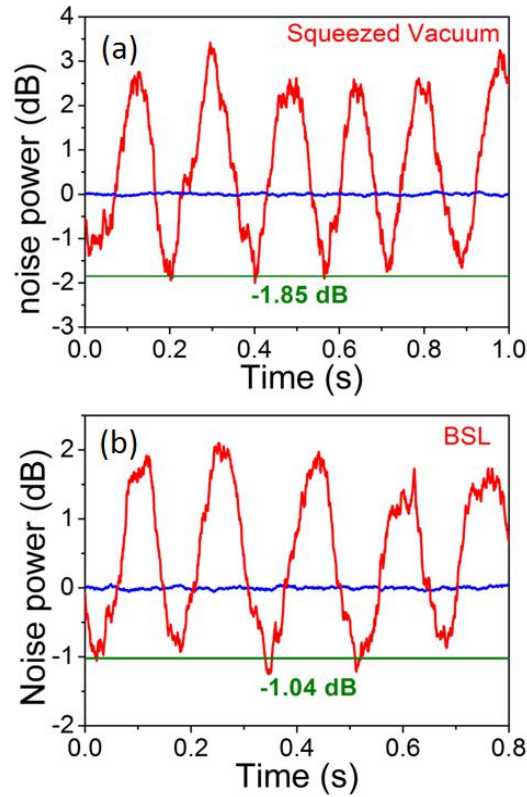

**Fig .4.** Normalized noise power (red) at 2.5 MHz as a function of the phase scanning time (with 3 Hz triangle-wave modulation). The analyzer resolution and the video bandwidths of the electronic spectrum analyzer are 100 kHz to 10 Hz, respectively. The quantum noise limit (blue) is measured with average when squeezing is blocked. (a) Squeezed vacuum without seed injection. (b) BSL with seed coherent state injection.

The optical spectrum of the generated squeezed vacuum and BSL are shown in Fig. 5. (a) and (b). Due to the strength of the squeezing vacuum is too weak to be measured directly with the optical spectrum analyzer (OSA), an indirect measurement method using photon counting is used by employing a tunable fiber filter with a 140 nm tunable width and an avalanche photodiode based single photon counter. With the 5 nm tuning step and the 1 nm filter bandwidth, the estimated spectral width is about 60 nm based on the curve in Fig.5 (a), which is determined by a phase match acceptance bandwidth of the nonlinear process. The spectrum of the BSL directly measured by an OSA is shown in Fig. 5. (b), it is exactly the same as that of the seed laser. With the self-heterodyne method involving a 48.8 km fiber delayed Mach-Zehnder interferometer, the linewidth of the 1550 nm seed laser is measured to be 30 kHz. It can be conservatively speculated that the linewidth of the BSL produced by the double-pass OPA is approximately equal to 30 kHz. This means that, unlike broadband squeezing vacuum, the narrowband characteristic of the BSL make it unrestricted in wavelength-dependent loss systems. Further, the narrowband characteristic will make it perform better in the field of precise measurement such as fiber sensing systems.

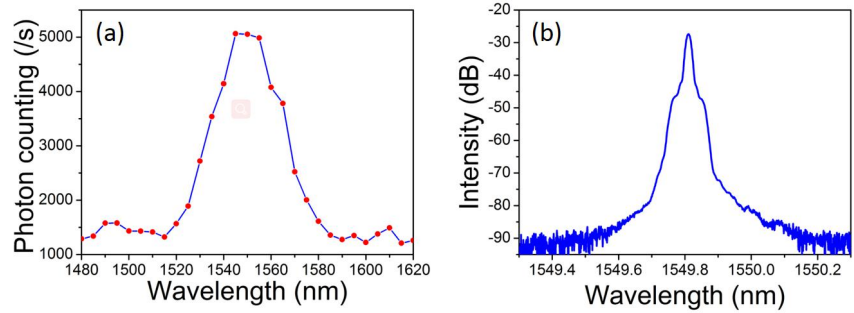

**Fig. 5.** (a) Spectrum of the squeezed vacuum estimated by counting the number of photons at different wavelengths. (b) Spectrum of the BSL measured by an optical spectrum analyzer with a resolution of 0.016nm.

In conclusion, a telecom fiber-based narrowband BSL generated by double-pass parametric amplification is developed. A squeezing value of 1.04 dB and a spectral bandwidth of 30 kHz are obtained. As far as we know, this is the first time that BSL has been reported in a fiber-based system to date. Although this BSL is still subject to large loss and phase jitter of the system now, with the advancement of materials and device technology, and the introduction of phase locking technology, this kind of BSL based on double-pass parametric amplification will likely be widely used in the field of quantum precision measurement and quantum information with its simple and fiber-based characteristics [34, 35].

**Funding Information.** National Key Research and Development Program of China (2017YFF0104602), Major Program of the National Natural Science Foundation of China (61790582), NSFC (61635004, 11674103, 61535014 and 51772101), Guangdong Key Research and Development Program (2018B090904001 and 2018B090904003), Local Innovative and Research Teams Project of Guangdong Pearl River Talents Program (2017BT01X137), Guangdong Natural Science Foundation (2017A030310007), the Science and Technology Project of Guangdong (2016B090925004 and 2017B090911005), and the Science and Technology Project of Guangzhou (201804020028).